\documentclass[
reprint,
superscriptaddress,
showpacs,
preprintnumbers,
amsmath,
amssymb,
aps,  
floatfix,
] {revtex4-2}

\usepackage{graphicx}  
\usepackage{dcolumn} 
\usepackage{bm}        
\usepackage{float}
\usepackage{siunitx}
\usepackage{xcolor}

\usepackage{graphicx}
\usepackage{upgreek}

\begin{document}

\author{M. R. Edwards}
\email{mredwards@stanford.edu}
\affiliation{Stanford University, Stanford, CA 94305}
\affiliation{Lawrence Livermore National Laboratory, Livermore, California, 94550}

\author{N. M. Fasano}
\author{A. M. Giakas}
\author{M. M. Wang}
\author{J. Griff-McMahon}
\author{A. Morozov}
\affiliation{Princeton University, Princeton, New Jersey, 08544}

\author{V. M. Perez-Ramirez}
\affiliation{Stanford University, Stanford, CA 94305}

\author{N. Lemos}
\author{P. Michel}
\affiliation{Lawrence Livermore National Laboratory, Livermore, California, 94550}

\author{J. M. Mikhailova}
\affiliation{Princeton University, Princeton, New Jersey, 08544}

\title{Greater than five-order-of-magnitude post-compression temporal contrast improvement with an ionization plasma grating}

\begin{abstract}
High-intensity lasers require suppression of prepulses and other non-ideal temporal structure to avoid target disruption before the arrival of the main pulse. To address this, we demonstrate that ionization gratings act as a controllable optical switch for high-power light with a temporal contrast improvement of at least $3\times10^5$ and a switching time less than 500 fs. We also show that a grating system can run for hours at 10 Hz without degradation. The contrast improvement from an ionization grating compares favorably to that achievable with plasma mirrors. 
\end{abstract}

\maketitle

High-power chirped-pulse-amplification (CPA) \cite{Strickland1985compression} lasers are the primary technology for creating extreme electric and magnetic fields; current systems produce femtosecond pulses focusable to light intensities above $10^{23}$ W/cm$^2$ \cite{Yoon2021realization}, well into the domain of relativistic optics \cite{Mourou2006optics}. 
The main peak of a high-power CPA pulse may be perfectly Gaussian, but the laser will also emit orders-of-magnitude less-intense light over several nanoseconds. At this scale the typical pulse structure is complex, with a picosecond pedestal from imperfect pulse compression, a nanosecond pedestal from amplified spontaneous emission, and prepulses: early-arriving copies of the main pulse.
For high-power lasers this premature light may contain substantial energy, sufficient to ionize, destroy, or disrupt targets well before the main-pulse interaction, which poses a particular problem for experiments on ion-acceleration \cite{Macchi2013ion}, relativistic high-order harmonic generation \cite{Dollar2013scaling,Edwards2020x}, and nanostructured surfaces \cite{Bargsten2017energy}.
Methods for improving temporal contrast (the intensity ratio of pre-arriving light to the main pulse) include cross-polarized wave generation \cite{Jullien200510$10$,Chvykov2006generation}, self-diffraction \cite{Liu2010temporal}, double CPA \cite{Kalashnikov2004characterization}, optical parametric amplification \cite{Shah2009high,Kiriyama2010high}, frequency doubling \cite{Marcinkevivcius2004frequency,Wang20170.85}, and saturable absorption \cite{Yu2012generation}, but the current standard for contrast improvement after the compressor is the plasma mirror \cite{Kapteyn1991prepulse,Backus1993prepulse,Thaury2007plasma,Mikhailova2011ultra}, where an anti-reflection (AR) coated solid surface is ionized to form an overdense (reflective) plasma, producing typical contrast increases of just over two orders of magnitude with up to 80\% efficiency for a single plasma mirror \cite{Roedel2011high,Scott2015optimization,Edwards2020multi} or four orders of magnitude with 50-60\% efficiency for a double mirror \cite{Kim2011spatio,Levy2007double}. The quality of the AR coating determines the contrast improvement; coating optimization leads to better suppression \cite{Inoue2016single}.
Specular reflection from a plasma mirror is in the same direction as that from the original surface, so the degree to which an AR coating can suppress the non-ionized reflectivity determines the achievable temporal contrast.
Although plasma mirrors have been demonstrated at 10 Hz and 1 kHz repetition rates \cite{Borot2014high}, in practice many systems, especially those using double-plasma mirrors, run in single-shot mode (below 1 Hz) due to the difficulty of rapidly replacing the target surface and the finite target lifetime.
The contrast improvement and repetition-rate limits of plasma mirrors suggest that future high-repetition-rate high-peak-power laser systems will require new approaches for fine control of temporal structure. 

Here we report the first experimental measurements of contrast for pulses diffracted from an ionization grating. The fundamental advantage of a plasma grating over a plasma mirror as a contrast improving optic is that before the arrival of the pump pulses there is no preferential optical path in what will be the direction of the diffracted beam; only non-directional scattering is present. We show that these gratings, which diffract rather than reflect light, are capable of improving contrast by at least five orders of magnitude and tolerate extended high-repetition-rate operation, making them potentially suitable as contrast-cleaning optics for femtosecond lasers. 

An ionization grating forms when two equal-wavelength laser pulses with intensity sufficient for ionization cross at a finite angle in a neutral gas, ionizing only the gas in their constructive interference fringes \cite{Suntsov2009femtosecond}. 
The plasma and residual neutral gas have different indices of refraction ($n$), so the modulated intensity interference pattern imprints a spatially varying refractive index: $n(x) = n_0 + \sum_{m=1}^\infty n_m \cos(2\pi m x / \Lambda - \phi_m)$, where we have expressed the periodic modulation as a sum of Fourier modes with phase $\phi_m$, and the fundamental period depends on the pump laser wavelength ($\lambda_0$) and crossing half-angle ($\theta_0$) as $\Lambda = \lambda_0 / 2 \sin \theta_0$. This spatial distribution of refractive index is a volume grating and can diffract a subsequent probe pulse (wavelength $\lambda_1$) incident at the Bragg angle [$\theta_B = \sin^{-1} \lambda_1/(2\Lambda)$].
Ionization gratings have previously been demonstrated primarily in atmospheric air \cite{Yang2010femtosecond,Liu2011two,Shi2011generation,Edwards2021measuring} or uniform gas \cite{Durand2012dynamics} environments, where filamentation, nonlinear propagation over extended distances, and the non-ideal shape of a grating defined by the overlap between two laser beams produced diffraction efficiencies (the fraction of incident probe light in the first order beam) between 0.1\% and 20\%, often with poor spatial quality. 
More recent experiments using gas jets in vacuum have reached average efficiencies above 35\% and single-shot efficiencies as high as 60\% \cite{Edwards2023control}, suggesting that ionization gratings can be sufficiently efficient for use as beamline optics. 
The temporal structure of pulses diffracted from ionization gratings have not previously been measured, in part because contrast measurements require at least 100-$\upmu$J-scale pulse energy, a clean spatial mode, and stable operation over hours, placing stringent requirements on grating quality and reproducability.

\begin{figure}[]
\centering
\includegraphics[width=\linewidth]{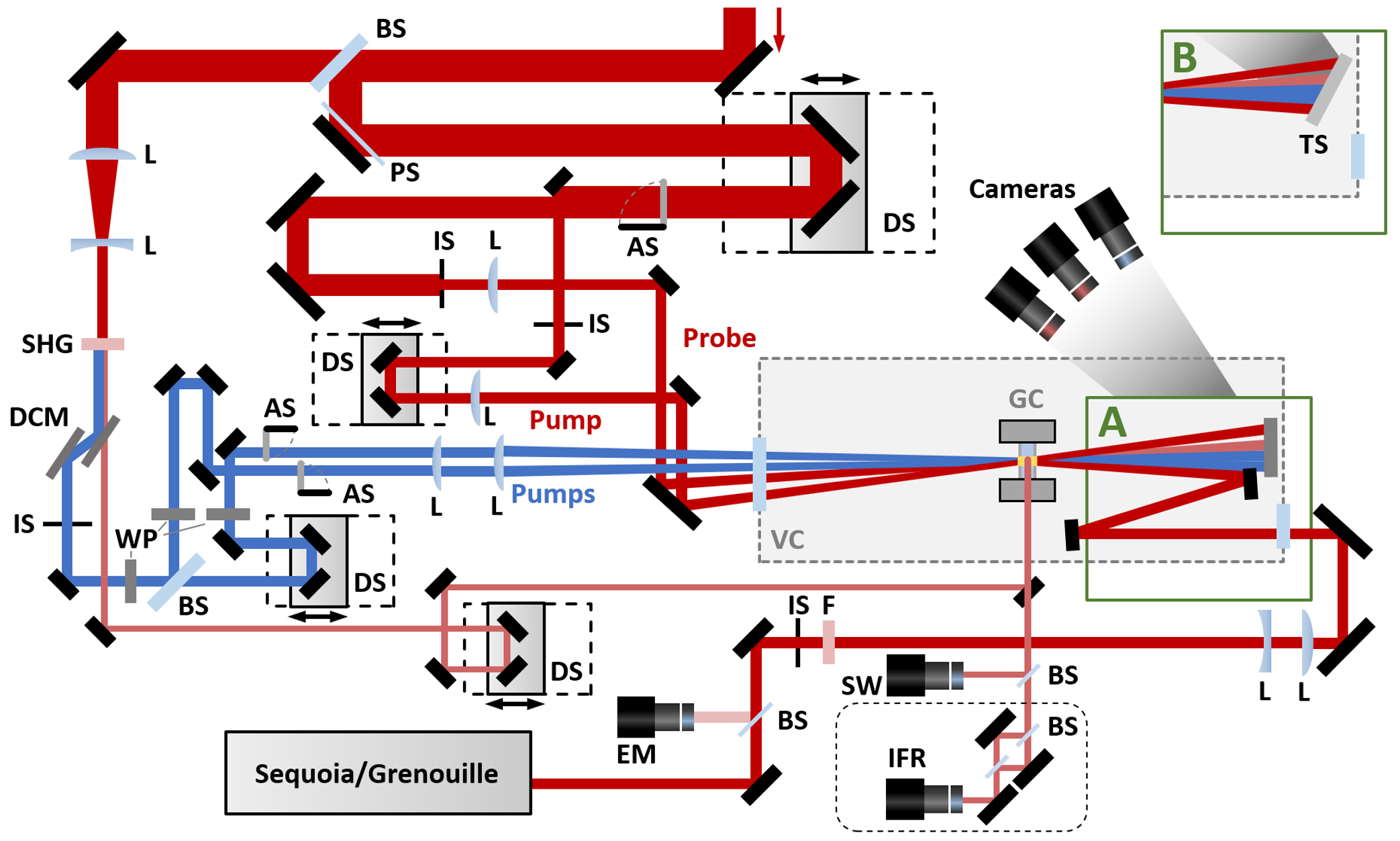}
\caption{Schematic of experiment, showing (A) configuration for measuring contrast and (B) configuration for quantifying efficiency and beam profiles. Key components are abbreviated as: BS: beam splitter, PS: prepulse slide, L: lens, IS: iris/aperture, AS: automatic shutter, DS: delay stage, SHG: second harmonic generation crystal (BBO: barium borate), DCM: dichroic mirror, WP: half waveplate, F: filter, EM: energy monitoring camera, IFR: interferometer, SW: shadowgraph/Schlieren camera, VC: vacuum chamber, GC: gas cell, TS: Teflon screen.}
\label{fig:schematic}
\end{figure}

To create ionization gratings with the requisite performance, we crossed two 400 nm pump pulses (0.15-0.38 mJ each, $\approx$ 35 fs full-width-half-maximum [FWHM] duration, $115 \:\upmu\textrm{m}\: [x] \times70 \:\upmu\textrm{m}\:[y]$ FWHM elliptical profile) at half-angle $\theta_0 = 0.7^\circ$ and a third (cotimed) pump at 800 nm and $2.3^\circ$ (1-3 mJ, $\approx$ 40 fs, $90\:\upmu\textrm{m}\:[x]\times70\:\upmu\textrm{m}\:[y]$ FWHM elliptical profile) in a gas cell inside a vacuum chamber, as drawn in Fig.~\ref{fig:schematic}. The cell was filled with CO$_2$ or air at between 100 and 2000 mbar and comprised two 50-$\upmu$m-thick steel sheets separated by a 1-mm gas-filled gap, an aluminum frame, and sapphire windows on the perpendicular faces providing optical access. The beams burned an entrance and an exit aperture through the steel sheets, providing an optical path through the cell. 
Although the standard ionization grating configuration uses only two pump beams, and we were able to observe diffraction without the additional 800-nm pump,  the third pump increased the plasma density and improved the peak diffraction efficiency by about an order of magnitude. Diffraction from this pump, intentionally not incident at the Bragg angle, was measured as at least three orders of magnitude weaker than the diffracted probe. 
The 800-nm probe pulse with up to 1.5 mJ (34 fs, $70 \:\upmu\textrm{m}$ FWHM circular profile) was incident at $\theta_1 = 1.4^\circ$ at a delay of 200 fs for the measurements presented here. 
Given the diameter and length of the grating and the probe angle of incidence, we can treat the configuration as a transmission grating with well-defined exit and entrance surfaces for the probe. 
%
We estimated the plasma density via interferometry with an 800-nm pulse perpendicular to the plane of the other beams. All five beams were derived from the same source and relative timing was controlled with delay stages. 
The individual beam spatial profiles were recorded with a camera placed at the gas cell position and we used an energy meter to record the pulse energy of each beam.
The diffraction efficiency was quantified by measuring the light of each beam scattered from a Teflon screen with a series of filtered cameras (Fig.~\ref{fig:schematic} configuration B) and was in good agreement with energy meter measurements.

\begin{figure}[]
\centering
\includegraphics[width=\linewidth]{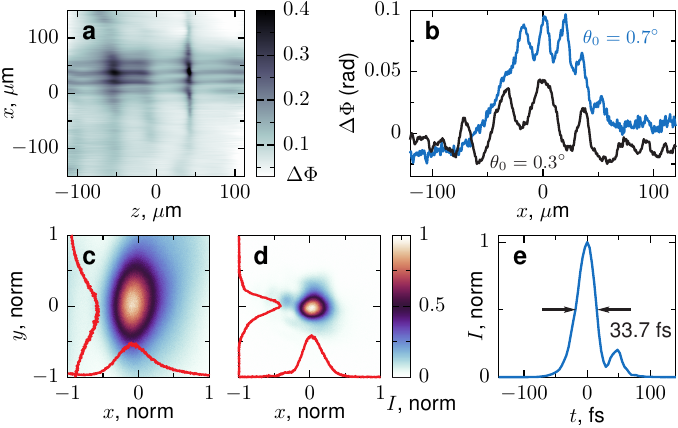}
\caption{Plasma grating density measurements via interferometry and beam profiles. (a) Two-dimensional map of phase shifts produced by grating showing horizontal fringes. Vertical marks are artifacts from incomplete subtraction of reflections from gas cell walls. (b) Line plot from grating shown in (a) at $\theta_0 = 0.7^\circ$ compared to the profile of a separate grating produced by pump beams with $\theta_0 = 0.3^\circ$. (c) Near-field diffracted beam profile measured using Teflon screen. (d) Far-field (focused) diffracted beam profile measured with beam focused on microscope objective and imaged with a camera. (e) FROG measurement of diffracted beam pulse shape.}
\label{fig:INFR}
\end{figure}

The optical properties of a volume grating derive from the spatial variation of the refractive index, especially the fundamental modulation amplitude $n_1$. 
Using a Mach-Zehnder interferometer, we directly measured a two-dimensional projection phase shift due to the distribution of refractive index within the grating, as shown in Fig.~\ref{fig:INFR}a where the grating structure (horizontal fringes) is clearly visible. Figure~\ref{fig:INFR}b shows the amplitude of the phase shift and the measured change of the fringe period with pump incidence angle, with the grating period closely agreeing with the expected angle-of-incidence dependence. 
The fringe structure removes the radial symmetry and precludes the standard Abel transform approach, so we estimate the plasma density using low-power beam profile measurements and the observed grating width (in $x$) to approximate the path length in $y$. 
This suggests a peak plasma density of $6\times10^{17}$ cm${^{-3}}$, which supports a maximum $n_1$ of approximately $4\times10^{-4}$. At the 1-1.5 mm effective length ($D$) of the grating (the density gradients at the entrance and exit apertures contribute to diffraction proportionally to the local $n_1$ \cite{Yeh1993introduction,Edwards2022plasma}), the maximum diffraction efficiency $\eta = \sin^2 [\pi n_1 D / (\lambda_1 \cos \theta_B)]$ \cite{Yeh1993introduction} is 1 (full diffraction, for $D = 1$ mm). However, the actual value of $n_1$ is likely lower than required for this maximum (the fringe spacing is not much larger than the spatial resolution of our interferometer, preventing precise extraction of the relative amplitudes of higher-order Fourier modes), the probe beam size is comparable to the grating size leading to edge effects, and the grating spectral [$\Delta \lambda_g \approx \lambda_0 n_1/(2\sin^2 \theta_B) \approx 250 \textrm{ nm}$] and angular [$\Delta \theta_g \approx n_1/\sin \theta_B \approx 0.9^\circ$] bandwidths \cite{Edwards2022plasma} are larger but not much larger than those of the probe ($80 \textrm{ nm}$, $0.5^\circ$, respectively), lowering maximum efficiency.
These effects lead to an average measured diffraction efficiency of 10\% for the grating in Fig.~\ref{fig:INFR}a, with some individual shots reaching 22\%. The beam profile images in Fig.~\ref{fig:INFR}cd show that the diffracted beam has high spatial quality in both the near and far field.
Pulse duration measurements made with a Grenouille implementation of frequency resolved optical grating (FROG) \cite{OShea2001highly} show that a short pulse duration (33.7 fs) and good temporal pulse shape is maintained in the diffracted beam (Fig.~\ref{fig:INFR}e).

Volume transmission gratings introduce angular dispersion and pulse front tilt. Angular dispersion is desirable for plasma gratings in a compressor \cite{Edwards2022plasma}, but unwanted for gratings meant for temporal cleaning without side effects. 
For small $\theta_B$, the spread in diffraction angles is $\Delta \theta = 2 \theta_B \Delta \lambda / \lambda_1$ \cite{Edwards2022plasma}. Here, at a probe incidence angle of 1.4$^\circ$, the expected difference in diffraction angle between the 760 nm and 830 nm edges of the probe spectrum is 0.25$^\circ$, smaller than the probe divergence angle. 
Both angular dispersion and pulse front tilt can be made arbitrarily small by decreasing the Bragg angle of the grating.

\begin{figure}[]
\centering
\includegraphics[width=\linewidth]{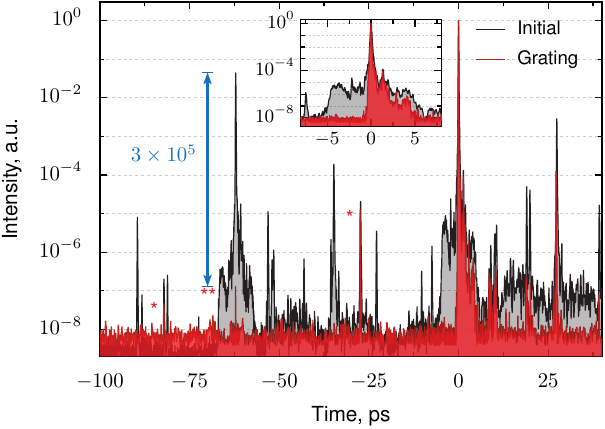}
\caption{Third-order autocorrelation measurement of temporal pulse shape with and without an ionization grating. Peaks marked (*) in the grating trace are measurement artifacts created by post-pulses. The residual signal (**) corresponds to an expected artifact from an observed post-pulse, and the real residual prepulse may be substantially weaker. The noise floor of the detector was at $10^{-8}$.}
\label{fig:contrast}
\end{figure}

We measured the contrast of the diffracted pulse using a third-order cross-correlator (Sequoia, Amplitude Laser) by replacing the Teflon screen with a mirror and recollimating optics and using apertures and filters to isolate the diffracted beam. The incident contrast of the laser was deliberately spoiled with a glass slide placed in front of a mirror (PS, Fig.~\ref{fig:schematic}), introducing a 4.5\% prepulse at -62 ps and a series of smaller prepulses (see the trace of the initial beam in Fig.~\ref{fig:contrast}), including postpulses of the new prepulse, prepulses caused by multiple reflections within the glass slide, and measurement artifacts from the strong prepulse inside the cross-correlator. Up to 10\% of the 1 mJ probe beam was used to measure the contrast trace of the diffracted beam, which is plotted as the grating trace in Fig.~\ref{fig:contrast}.
Diffraction from the grating removed almost all this prepulse structure from the beam: only three signatures are visible above the $10^{-8}$ noise floor of the detector at times earlier than -500 fs. The grating also entirely removed the picosecond-scale pedestal, resulting in an extremely clean rise of the main pulse over eight orders of magnitude. 
The two peaks (*) at -82 ps and -27 ps correspond to strong post-pulses and are likely ghost artifacts; the +27 ps post-pulse is introduced by a back-surface reflection from the first beamsplitter in the experimental system. 
The residual peak at -62 ps (**), $3\times 10^5$ times weaker than the original prepulse, sets the measured contrast improvement of the grating. However, since the glass side for the prepulse also introduced a corresponding post-pulse with sufficient amplitude to create a ghost artifact of this size, the actual contrast improvement of the grating may be somewhat higher. We observe non-directional scattered light from the probe in the gas cell at a level of $10^{-6}$, which may set a limit on contrast improvement, albeit one which is likely improvable with a more carefully configured experiment.
Similar performance was found over the range of pump and probe energies described above. This specific trace was collected with 400-nm pump energies of 0.28 mJ, an 800-nm pump energy of 3 mJ, 1.5 mJ in the probe beam, and a gas cell filled with air. 

\begin{figure}[]
\centering
\includegraphics[width=\linewidth]{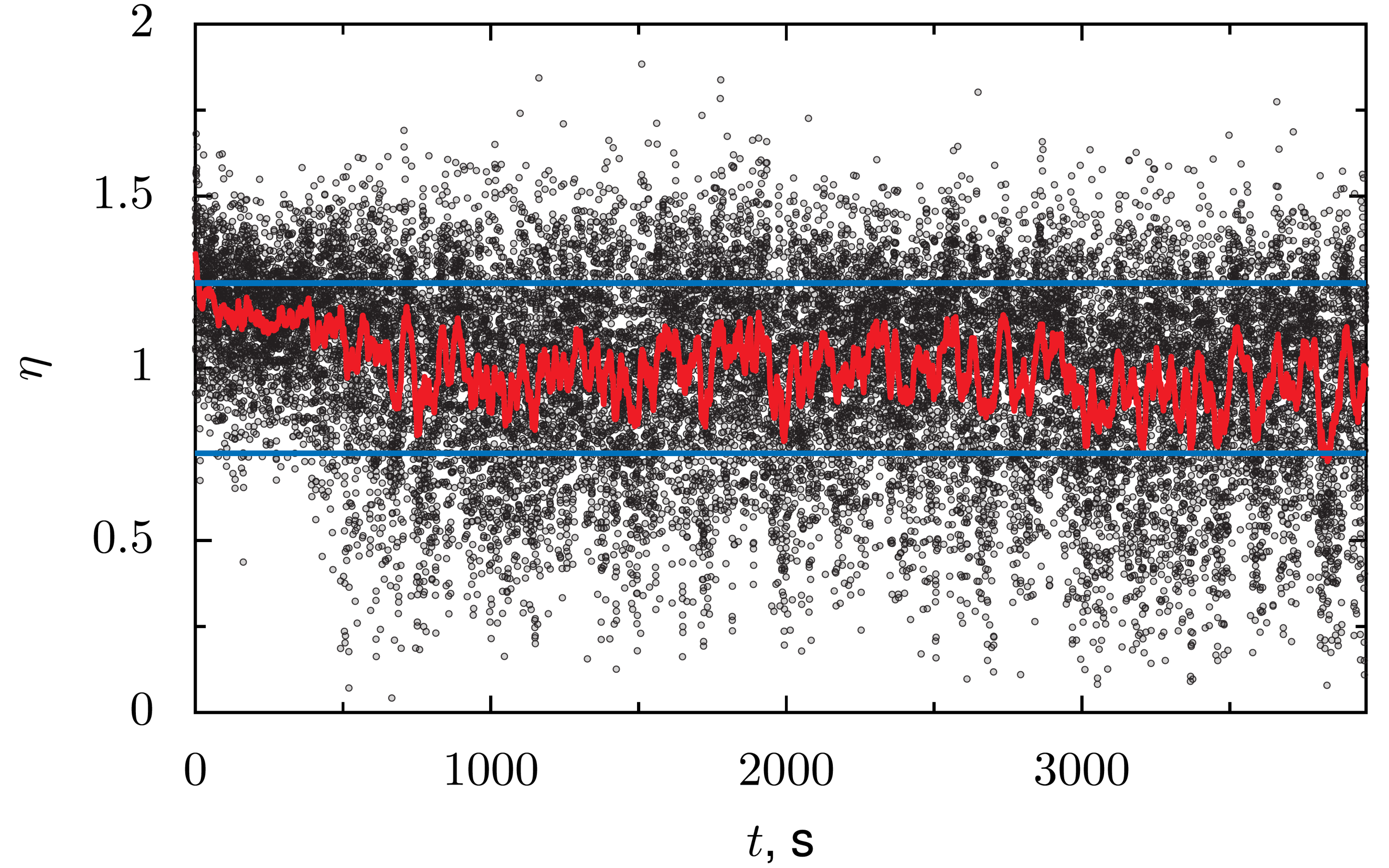}
\caption{Single-shot measurements of grating diffraction efficiency (normalized to mean). The system was run at 10 Hz, with shots recorded at 5 Hz. The plot shows just over an hour of operation. The red line indicates a 100-point moving average, and the blue lines show the standard deviation of the full measurement set.}
\label{fig:stability}
\end{figure}

The extreme contrast improvement we observe is possible due to the high-order nonlinearity of ionization; however, this makes the system sensitive to laser energy variation. As shown by Fig.~\ref{fig:stability}, shot-to-shot changes of the diffracted beam energy are substantial: 3\% energy fluctuations of the input laser pulse (standard deviation) become 25\% diffracted energy fluctuations, a level marked by blue lines. Despite this, the long-term average stability of the system is promising, and a 100-point average (red line) shows a much narrower range of fluctuations over an hour of almost constant average performance. The system was continuously operated for hours at a time without measurable performance degradation. The instability of the system appears to largely result from the nonlinearity of the second harmonic generation for the pumps and the ionization process, and to a more limited extent by vibrations affecting the beam profile overlaps. In addition to using a more stable initial beam, it may be possible to improve the shot-to-shot stability of the system by operating in a regime closer to saturation of the plasma density.  

The results we present here are promising, but the demonstrated efficiency and total energy capacity are too low to be immediately useful in a high-power beamline. Theory suggests that both higher efficiency diffraction and greater temporal contrast improvement are plausible, providing a path to higher-energy-capacity gratings for full-power terawatt and petawatt beams.
An ideal volume transmission grating in the Bragg regime ($\rho = \lambda_1^2/\Lambda^2 n_0 n_1 \gg 1$ \cite{Moharam1978criterion}) has a maximum theoretical diffraction efficiency of 100\%, so further experimental optimization may be able to further increase diffraction efficiency. A full accounting for efficiency must also include the energy in the pump pulses that create the grating, and to make the measurement shown in Fig.~\ref{fig:schematic} we used more energy in the pumps all together (3.6 mJ) than in the probe (1.5 mJ). Reducing the overall 
energy use may be practical with more efficient grating creation, including approaches like resonant multiphoton ionization.
The mechanisms limiting contrast improvement are still unclear. Other potential sources of pump-induced index modulation (e.g. neutral gas electronic and rotational nonlinearities \cite{Chen2007single}) produce much weaker modulation and also decrease quickly with pump intensity. Furthermore, the diffraction efficiency of weak gratings ($\eta \ll 1$) falls off as $n_1^2$, so all but the strongest ($>10\%$) prepulses are unlikely to produce measurable diffraction via these mechanisms. Non-directional light scattering depends on the exact geometry of the initial gas, and although it is difficult to estimate the degree to which it can be suppressed, we anticipate that contrast improvements much greater than $10^5$ are possible.
%
%
Reaching higher energy capacity requires increasing the grating diameter or choosing a less easily ionized initial gas, like helium; both routes are accessible with higher energy pumps.

We have considered an isolated grating deployed solely for contrast improvement, but ionization optics are also potential replacements for gratings or lenses in compact high-power CPA lasers \cite{Edwards2022holographic,Edwards2022plasma}. These measurements suggest that high contrast is a likely side effect of ionization-based diffractive optics. 
Plasma gratings based on other mechanisms, like the ponderomotive force \cite{Lehmann2016transient,Peng2019nonlinear}, could also improve contrast for longer-than-picosecond timescales, depending on the nonlinearity of the formation mechanism; good temporal contrast may be a general property of diffractive plasma-optic systems.

In summary, we have shown that the temporal contrast cleaning capability of ionization gratings compares well to that of plasma mirrors, with a higher degree of prepulse suppression than double plasma mirror systems and the ability to operate continuously at 10 Hz without replacing targets. High-repetition-rate continuous operation permitted a high-temporal-resolution measurement of the contrast of a diffracted pulse. With further work to improve energy capacity and diffraction efficiency, plasma gratings may prove to be an attractive method for contrast cleaning high-power femtosecond pulses. 

\medskip \noindent

\begin{acknowledgments}
Lawrence Livermore National Laboratory (20-ERD-057, 21-LW-013). U.S. DOE NNSA (DE-NA0004130). U.S. NSF (PHY 1806911, PHY 2206711, PHY 2308641). The Gordon and Betty Moore Foundation, GBMF12255, grant DOI 10.37807/gbmf12255. NMF and AMG acknowledge support from the Program in Plasma Science and Technology (PPST). Lawrence Livermore National Laboratory is operated by Lawrence Livermore National Security, LLC, for the U.S. Department of Energy, National Nuclear Security Administration under Contract DE-AC52-07NA27344.
\end{acknowledgments}




\bibliography{References}


\end{document}